\documentclass[aps,prb,twocolumn,showpacs,groupedaddress]{revtex4}

\usepackage{amssymb,amsmath}
\usepackage{graphicx}
\usepackage{dcolumn}
\usepackage{bm}

\begin{document}

\title{\emph{Ab initio} study of compressed Ar(H$_{2}$)$_{2}$: structural stability and anomalous melting }

\author{Claudio Cazorla$^{\rm a}$ and Daniel Errandonea$^{\rm b}$}
\affiliation{$^{\rm a}$ Department of Chemistry, University College London, London WC1H 0AJ, United Kingdom \\
             $^{\rm b}$ Departamento de F\'isica Aplicada -ICMUV-, Fundaci\'o General Universitat de Valencia, Spain}
\email{c.silva@ucl.ac.uk}

\begin{abstract}

We study the structural stability and dynamical properties of 
Ar(H$_{2}$)$_{2}$ under pressure using first-principles and \emph{ab initio} molecular dynamics techniques.
At low temperatures, Ar(H$_{2}$)$_{2}$ is found to stabilize in the cubic C15 Laves 
structure (MgCu$_{2}$) and not in the hexagonal C14 Laves structure (MgZn$_{2}$) 
as it has been assumed previously. Based on enthalpy energy and phonon 
calculations, we propose a temperature-induced MgCu$_{2}$~$\to$~MgZn$_{2}$ phase 
transition that may rationalize the existing discrepancies between the sets of 
Raman and infrared vibron measurements. Our AIMD simulations suggest 
that the melting line of Ar(H$_{2}$)$_{2}$ presents negative slope in 
the interval $60 \le P \le 110$~GPa. We explain the origin of this 
intriguing physical phenomenon in terms of decoupling of the Ar and H$_{2}$ 
degrees of freedom and effective thermal-like excitations arising from 
coexisting liquid H$_{2}$ and solid Ar phases. 

\end{abstract}


\maketitle

\section{Introduction}
\label{sec:intro}

Solid hydrogen H$_{2}$ is expected to become metallic at compressions higher  
than $400$~GPa~[\onlinecite{stadele00}]. In fact, experimental signatures of the 
long-sought insulator-to-metal phase transition remain elusive 
up to pressures of~$\sim 340$~GPa~[\onlinecite{narayana98}]. 
Accepted pressure-induced mechanisms by which the metallicity of hydrogen 
can be enhanced involve atomization of H$_{2}$ molecules and partial 
filling of electronic $\sigma_{u}^{*}$ molecular levels due to charge transfer 
from or band hybridization with other chemical 
species~[\onlinecite{loubeyre93,strobel09,zurek09}].
Already in the earlier 90's Loubeyre \emph{et al.}, based on the disappearance of the Raman (R) vibron mode
and the darkening of the material, claimed to observe  
metallization of the Ar(H$_{2}$)$_{2}$ compound when compressed 
in the diamond-anvil-cell (DAC) up to ~$\sim 200$~GPa~[\onlinecite{loubeyre93}]. 
The stable room-temperature (RT) phase structure of this 
compound was identified with the hexagonal C14 Laves structure typified by the MgZn$_{2}$ 
crystal (space group:~$P63/mmc$). 
Strikingly, posterior synchrotron infrared (IR) measurements did not show evidence of molecular bonding instabilities nor 
metallic Drude-like behavior up to at least $220$~GPa~[\onlinecite{datchi96}]. 
Subsequently, Bernard \emph{et al.} suggested that activation of H$_{2}$ dissociation 
processes and corresponding development of metallization in Ar(H$_{2}$)$_{2}$ could occur 
via a solid-solid phase transition of the 
MgZn$_{2}$~$\to$~AlB$_{2}$ (space group:~$P63/mmc$)
type at pressures already within the reach of DAC 
capabilities~[\onlinecite{bernard97}]. 
However, recent \emph{ab initio} work done by Matsumoto \emph{et al.} 
demonstrates that the onset of metallicity in the AlB$_{2}$ structure 
commences at pressures significantly higher than in  
pure bulk H$_{2}$~[\onlinecite{matsumoto07}]. 
 
In view of the growing interest on hydrogen-rich van der Waals (vdW) compounds under 
pressure~[\onlinecite{strobel09,somayazulu09}], partly motivated by the hydrogen-storage problem, and 
of the unresolved discrepancies described above, we have conducted a theoretical 
study on Ar(H$_{2}$)$_{2}$ under extreme $P - T$ conditions using first-principles 
density functional theory (DFT) calculations and \emph{ab initio} molecular dynamics 
simulations (AIMD). 
In this letter, we present results showing that at low temperatures
and pressures up to $\sim 215$~GPa the Ar(H$_{2}$)$_{2}$ crystal stabilizes in the cubic 
C15 Laves structure typified by the MgCu$_{2}$ solid (space group:~$Fd3m$). 
This structure has not been considered in previous  
works~[\onlinecite{bernard97,matsumoto07,chacham95}]
though its probable relevance to Ar(H$_{2}$)$_{2}$ was pointed out recently~[\onlinecite{cazorla09}]. 
On the light of first-principles enthalpy and phonon calculations, we propose a   
temperature-induced (pressure-induced) phase transition of the 
MgCu$_{2}$~$\to$~MgZn$_{2}$ (MgZn$_{2}$~$\to$~MgCu$_{2}$) type that 
may clarify the origin of the discrepancies between the sets of R and IR data.  
Furthermore, in the high-$P$ regime ($P \sim 210$~GPa) we find that a metallic 
hydrogen-rich liquid can be stabilized at temperatures of $\sim 4000$~K wherein H-H 
coordination features render molecular dissociation activity.
By means of AIMD simulations, we estimated an upper bound of the melting curve $P(T_{m})$ of 
Ar(H$_{2}$)$_{2}$ and found a negative $\partial T_{m} / \partial P$ slope
spanning over the interval $60 \le P \le 110$~GPa.
Our simulations show that the lattice composed by H$_{2}$ molecules melts at 
temperatures significantly lower than the lattice of Ar atoms does, so leading to stable
mixtures of coexisting liquid~H$_{2}$ and solid~Ar over wide $P - T$ ranges. 
We propose an argument based on this atypical physical behavior to 
explain the cause of the estimated negative $\partial T_{m} / \partial P$ slope.

\begin{figure}
\centerline{
        \includegraphics[width=1.00\linewidth,angle=0]{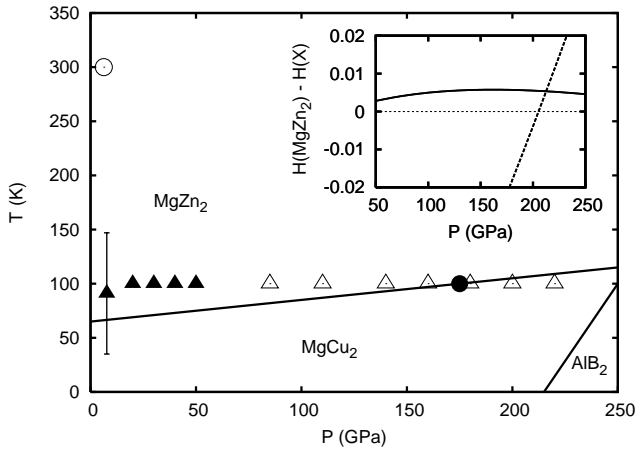}}%
\vspace{-0.25cm}
        \caption{Schematic low-$T$ phase diagram of Ar(H$_{2}$)$_{2}$
                 under pressure. MgZn$_{2}$~-~MgCu$_{2}$ and MgCu$_{2}$~-~AlB$_{2}$
                 phase boundaries are sketched according to the  
                 results and arguments presented in the text.   
                 Thermodynamic states at which x-ray, R and IR vibron measurements  
                 were carried out are indicated:  
		 [\onlinecite{loubeyre93}]~x-ray = $\circ$,
		 [\onlinecite{loubeyre93}]~R = $\bullet$,
                 ~[\onlinecite{ulivi99}]~IR = $\blacktriangle$
                 and~[\onlinecite{datchi96}]~IR = $\vartriangle$.
                 \emph{Inset}: Enthalpy difference per particle of the MgCu$_{2}$
                 (solid line) and AlB$_{2}$ (dashed line) structures with respect to 
                 the MgZn$_{2}$ Laves phase as function of pressure at zero temperature.}
\label{fig:PD-H}
\end{figure}

\section{Overview of the calculations}
\label{sec:method}

Our calculations were performed using the all-electron projector augmented
wave method and generalized gradient approximation of 
Wang and Perdew as implemented in the VASP 
code~[\onlinecite{vasp}]. Dense Monkhorst-Pack special 
$k$-point meshes~[\onlinecite{monkhorst76}] for sampling of the 
first Brillouin zone (IBZ) and a cutoff energy of $400$~eV were employed 
to guarantee convergence of the total energy per particle 
to within $0.5$~meV. In particular, we used $8\times 8\times 8$,
$8\times 8\times 4$ and $12\times 12\times 12$ 
$k$-point grids for calculations on the perfect unit cell corresponding to the 
MgCu$_{2}$, MgZn$_{2}$ and AlB$_{2}$ crystal structures, respectively.  
All the considered crystal structures were relaxed using a conjugate-gradient algorithm 
and imposing the forces on the particles to be less than $0.005$~eV/\AA.   
The phonon frequencies in our calculations were obtained using
the small-displacement method~[\onlinecite{kresse95,phon}] over the unit
cells ($\Gamma$-point phonon frequencies) and large supercells 
containing $80$ atoms.
\emph{Ab initio} molecular dynamics simulations were carried out in the 
canonical ensemble $(N , V , T)$ using bulk supercells of Ar(H$_{2}$)$_{2}$ containing $160$
atoms ($\Gamma$-point sampling). At given pressure, the dynamical properties of the 
system were sampled at $400$~K intervals from zero-temperature up to the  
melting curve of pure Ar. Temperatures were maintained using 
Nos\'e-Hoover thermostats. A typical AIMD simulation consisted of $3$~ps 
of thermalization followed by $7$~ps over which statistical averages were taken.    
It is worth noticing that we recently used a very similar computational approach 
to the one described here to study the energetic and structural properties
of Ar(He)$_{2}$ and Ne(He)$_{2}$ crystals under pressure~[\onlinecite{cazorla09}], 
and that very recent experiments~[\onlinecite{fukui10}] have 
confirmed the validity of our predicted $P(V)$ curves.  

\begin{figure}
\centerline{
        \includegraphics[width=1.00\linewidth,angle=0]{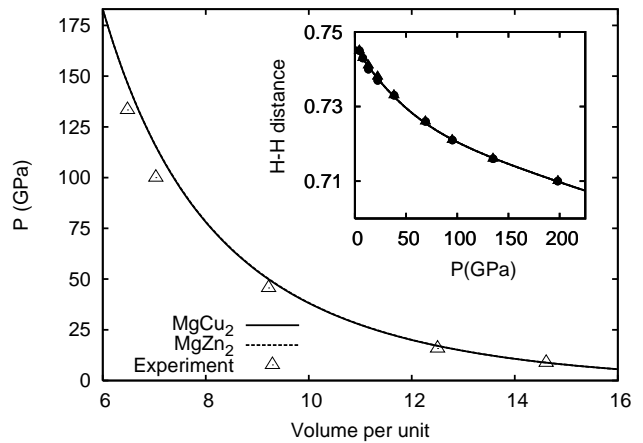}}%
\vspace{-0.25cm}
        \caption{Calculated zero-temperature equation of state of Ar(H$_{2}$)$_{2}$
                 in the MgCu$_{2}$ (solid line) and MgZn$_{2}$ (dashed
                 line) crystal Laves structures. 
                 Experimental data is from work~[\onlinecite{bernard97}].\emph{Inset}:
                 calculated pressure-dependence of the H-H intermolecular 
                 distance in the MgCu$_{2}$ (solid line, solid
                 dots) and MgZn$_{2}$ (dashed line, solid triangles)
                 crystal structures. Volumes and distances are in units
                 of \AA$^{3}$/particle and \AA, respectively.}
\label{fig:EOS}
\end{figure}

\section{Results and Discussion}
\label{sec:results}

\subsection{Low-$T$ results}
\label{subsec:lowT}

A series of candidate structures
were considered in our \emph{ab initio} enthalpy $H ( P )$ calculations
(rutile, fluorite, MgNi$_{2}$, etc.) however 
only the MgZn$_{2}$, MgCu$_{2}$ and AlB$_{2}$ structures turned  
out to be energetically competitive so the following analysis 
concentrates on these ones.  
Ignoring quantum zero-point motion effects, we found that 
Ar(H$_{2}$)$_{2}$ is energetically more stable in the MgCu$_{2}$ structure 
than in either the MgZn$_{2}$ or AlB$_{2}$ structures
at pressures $P \le 215$~GPa (see Figure~\ref{fig:PD-H}). 
Also we predicted that at $P_{t} = 215(1)$~GPa the 
solid transitates from the MgCu$_{2}$ to the AlB$_{2}$  
structure. The calculated zero-temperature equation of state of 
Ar(H$_{2}$)$_{2}$ in the MgCu$_{2}$ structure displays very good agreement
with respect to the available experimental data~[\onlinecite{bernard97}]
(see Figure~\ref{fig:EOS}). For instance, at $V = 14.61$ and 
$12.50$~\AA$^{3}$/particle
we obtain a pressure of $8.83$ and $17.14$~GPa, respectively, 
to be compared with the experimental values $8.77$ and $15.79$~GPa.
It must be noted that the estimated equation of state and 
pressure-dependence of the H-H intermolecular bond distance of the
MgZn$_{2}$ and MgCu$_{2}$ phases appear to be indistinguishable 
within the numerical uncertainty in our calculations
(namely, $1$~GPa and $0.01$~\AA -see Figure~\ref{fig:EOS}-).  
 
\begin{figure}
\centerline{
        \includegraphics[width=1.00\linewidth,angle=0]{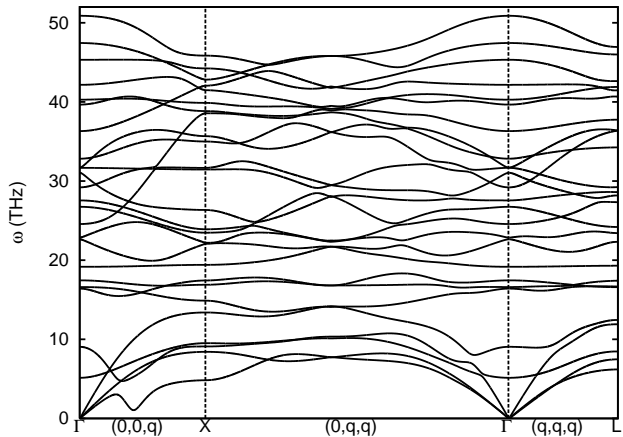}}%
\vspace{-0.25cm}	
        \caption{Calculated phonon frequency spectra of Ar(H$_{2}$)$_{2}$
                in the MgCu$_{2}$ crystal structure at $V = 5.83$~\AA$^{3}$/particle
                and $P \sim 200$~GPa (corresponding molecular vibron frequencies are not shown).      
                The calculation was done on a supercell containing $16$ Ar and $64$ H atoms
                and using a $4\times 4\times 4$ $k$-point grid for IBZ sampling.} 
\label{fig:phonons}
\end{figure}

Since hydrogen is a very light molecule, quantum 
zero-point motion corrections must be included in the calculations. 
Customarily this is achieved using quasi-harmonic 
approaches that involve estimation of the vibrational phonon 
frequencies. 
In following this procedure, we found that Ar(H$_{2}$)$_{2}$ in 
the MgZn$_{2}$ structure always exhibits 
imaginary $\Gamma$-phonon frequencies at pressures below 
$\sim 300$~GPa. 
This means that the MgZn$_{2}$ structure is mechanically
unstable, at least, at low temperatures. 
On the other hand, we found that the MgCu$_{2}$ structure is 
perfectly stable at all the studied pressures. 
Our numerical tests showed that this result does not depend on the 
approximation of the exchange-correlation functional 
used~[\onlinecite{ceperley80}]. 
In Figure~\ref{fig:phonons}, we show the phonon frequency spectra
of Ar(H$_{2}$)$_{2}$ in the MgCu$_{2}$ crystal structure calculated
along a $k$-point path contained within the IBZ and    
at pressure $\sim 200$~GPa .

Since the x-ray resolved RT crystal structure of 
Ar(H$_{2}$)$_{2}$ is MgZn$_{2}$~[\onlinecite{loubeyre93}],
we carried out a series of AIMD simulations of this and the MgCu$_{2}$ 
structure at $T = 300$~K in order to explore their stability.  
In fact, analysis of our simulations based on estimation of the  
averaged mean-squared displacement and position-correlation 
function~[\onlinecite{vocadlo03}] showed that both MgCu$_{2}$ and MgZn$_{2}$ 
structures are mechanically stable at RT. 
On view of these results and the reported x-ray data, we concluded that a 
temperature-induced MgCu$_{2}$~$\to$~MgZn$_{2}$ transition occurs at 
fixed $P$ (or equivalently, a pressure-induced MgZn$_{2}$~$\to$~MgCu$_{2}$ 
transition at fixed $T$ - see Figure~\ref{fig:PD-H} -). 
This temperature-induced transformation can be understood in terms of entropy: 
Ar(H$_{2}$)$_{2}$ in the MgZn$_{2}$ structure is highly anharmonic so ionic  
entropy contributions to the total free-energy stabilize this
structure over the MgCu$_{2}$ phase with raising temperature.          
In fact, IR measurements displayed an splitting of the vibron mode in the  
interval $35 \le T \le 150$~K ~[\onlinecite{ulivi99}]. 
This sppliting can be induced by the degenerancy removal of the infrarred
${\rm E_{1u}}$ mode due to the decrease of the crystal quality caused by precursor 
effects of the MgZn$_{2}$~$\to$~MgCu$_{2}$ transition.

\begin{figure}
\centerline{
        \includegraphics[width=1.00\linewidth,angle=0]{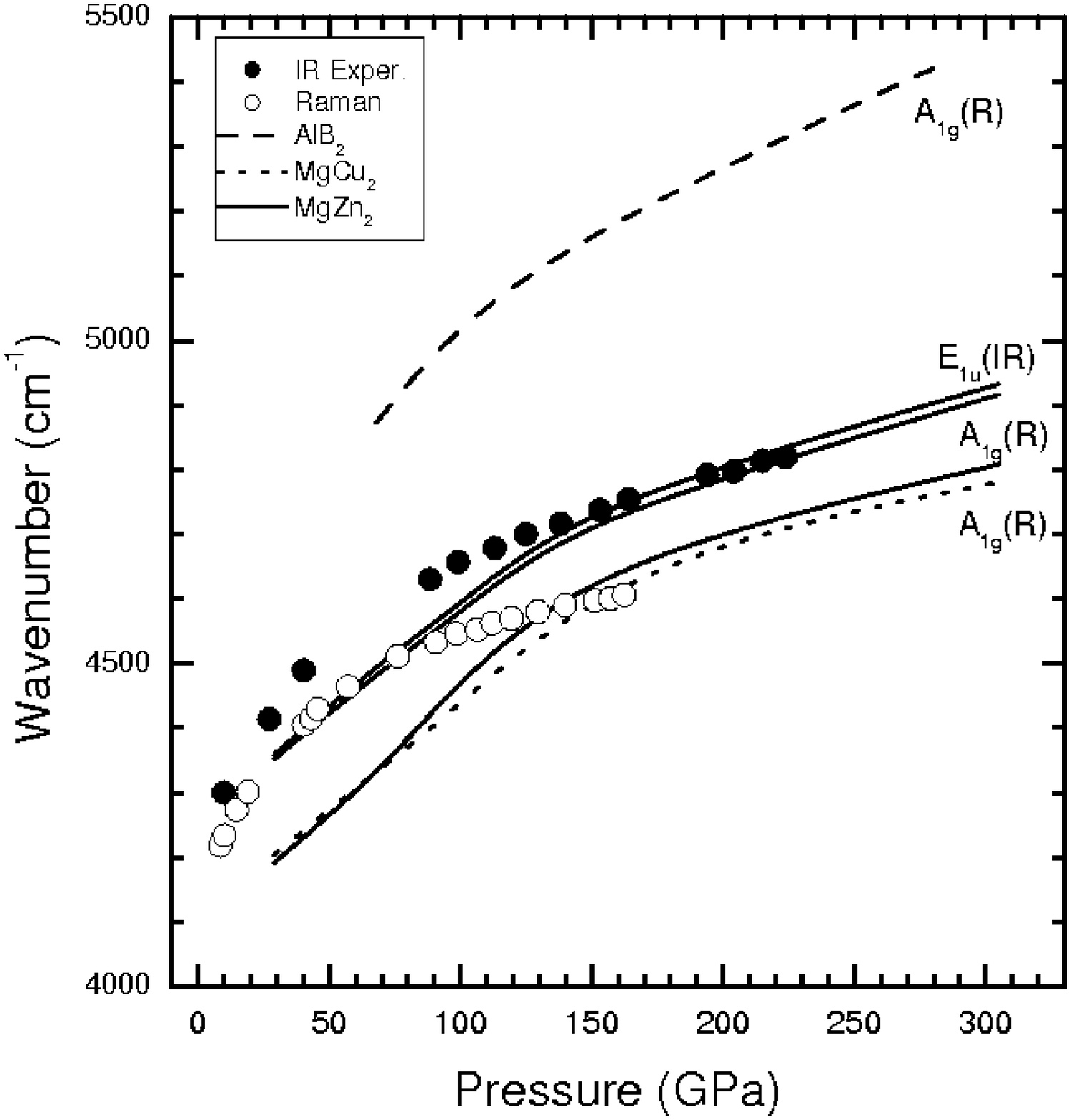}}%
\vspace{-1.25cm}	
        \caption{Calculated R and IR vibron lines of Ar(H$_{2}$)$_{2}$
                 in the MgZn$_{2}$, MgCu$_{2}$ and AlB$_{2}$ structures
                 as function of pressure. Experimental data from 
                 [\onlinecite{loubeyre93}]~($\circ$) and 
                 [\onlinecite{datchi96}]~($\bullet$) 
                 are shown.}      
\label{fig:vibron}
\end{figure}

The causes of the aforementioned R-IR experimental disagreements can be 
rationalized in the light of the foreseen MgCu$_{2}$~$\to$~MgZn$_{2}$  
transition, as we explain in what follows.
In Figure~\ref{fig:vibron}, we enclose the calculated R and IR vibron 
frequencies of Ar(H$_{2}$)$_{2}$ in the MgZn$_{2}$, MgCu$_{2}$ and AlB$_{2}$ 
structures as function of pressure. At high pressures, it is shown that 
the ${\rm A_{1g}~(R)}$ vibron line corresponding to the MgZn$_{2}$ and 
MgCu$_{2}$ structures are practically identical, whereas they settle 
appreciably below the line estimated for the AlB$_{2}$ structure, 
and follow closely the low-$T$ R results ~[\onlinecite{loubeyre93}].
These outcomes together with the energy and phonon results already presented, 
led us to think that the sudden dissapearance of the R vibron mode   
observed at high-$P$ might be related to the MgCu$_{2}$~$\to$~MgZn$_{2}$ 
transition unravelled in this Letter and not to pressure-induced 
metallic-like behavior~[\onlinecite{loubeyre93}] or MgZn$_{2}$~$\to$~AlB$_{2}$ 
phase transition~[\onlinecite{bernard97}] as previously suggested. 
In fact, accurate electronic density of states (DOS) analysis performed on  
the perfect MgZn$_{2}$, MgCu$_{2}$ and AlB$_{2}$ crystal structures 
show no closure of the electronic band gap up to compressions of at least $420$~GPa. 
Furthermore, the ${\rm E_{1u}~(IR)}$ vibron 
line that we calculated for the MgZn$_{2}$ structure (see Figure~\ref{fig:vibron}) 
agrees closely with Datchi's IR data obtained at high pressures.
This accordance is coherent if one considers that the series of IR  
experiments were realized near the edge of the MgCu$_{2}$-MgZn$_{2}$  
phase boundary, as we sketch in Figure~\ref{fig:PD-H}.
Conclusive experiments confirming the validity of our statements
might consist of new series of x-ray and IR measurements performed at 
temperatures below $100$~K and high pressures; 
according to our predictions, the vibron 
signature will eventually disappear at the crossing with 
the MgCu$_{2}$-MgZn$_{2}$ phase boundary since IR vibron modes
in the MgCu$_{2}$ structure are inactive.

\begin{figure}
\centerline{
        \includegraphics[width=1.00\linewidth,angle=0]{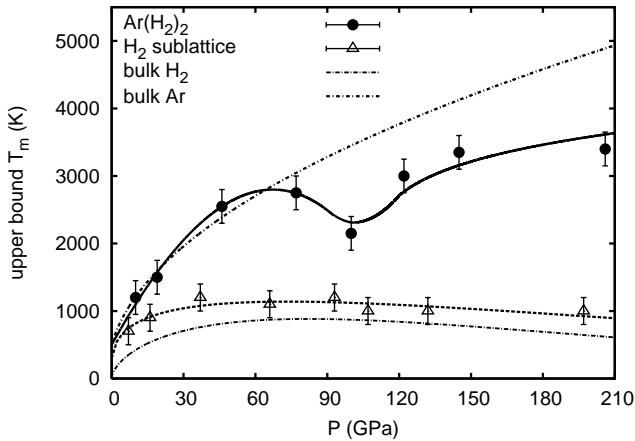}}%
\vspace{-0.25cm}
        \caption{Upper bound of the melting line of Ar(H$_{2}$)$_{2}$
                 under pressure (solid line). Melting states
                 obtained in the AIMD simulations are indicated
                 with $\bullet$ and corresponding error bars amount to $200$~K. 
                 Thermodynamic states at which the lattice of  
                 hydrogen molecules is observed to melt in the simulations are 
                 represented by $\vartriangle$ (fitted to a dashed line). 
                 The melting line of pure Ar~[\onlinecite{pechenik08}] 
                 (long-dashed and dotted line) and H$_{2}$~[\onlinecite{bonev04}] 
                 (dashed and dotted line) are shown for comparison.} 
\label{fig:ub-melting}
\end{figure}

\subsection{High-$T$ results}
\label{subsec:highT}

An intriguing physical phenomenon 
has been recently predicted and subsequently observed in H$_{2}$ under 
pressure. It consists in the appearance of a maximum peak on its melting line 
followed by a negative $\partial T_{m} / \partial P$ 
slope~[\onlinecite{bonev04,deemyad08}]. 
This phenomenon have been explained in terms of 
subtle changes on the intermolecular forces due to 
compression instead of more familiar arguments like 
promotion of valence electrons to higher energy orbitals or
occurrence of molecular dissociation processes.
Motivated by these interesting findings on hydrogen, we investigated 
the dynamical properties of Ar(H$_{2}$)$_{2}$ at high-$P$ and
high-$T$ in the search of similar physical manifestations and  
to provide further understanding of vdW compounds in general. 
There exist several well-established techniques by which 
one can determine the melting line of a material; these essentially 
base on solid-liquid phase coexistence simulations and/or Gibbs free-energy 
calculations~[\onlinecite{gillan}].      
Application of these methods at quantum first-principles 
level of description, however, turns out to be computationally very intensive
and laborious.
A simulation approach that has proved successful in 
reproducing general melting trends in materials at 
affordable computational cost, including 
experimental negative melting slopes, is the 
`heat-until-it-melts' method~[\onlinecite{tamblyn08,raty07}]. 
This technique allows for estimation of a precise upper bound of the  
solid-liquid phase boundary of interest. It must be stressed 
that we did not pursue accurate calculation of the melting line of 
Ar(H$_{2}$)$_{2}$ but to identify possible anomalous effects on it. 

\begin{figure}
\centerline{
        \includegraphics[width=1.00\linewidth,angle=0]{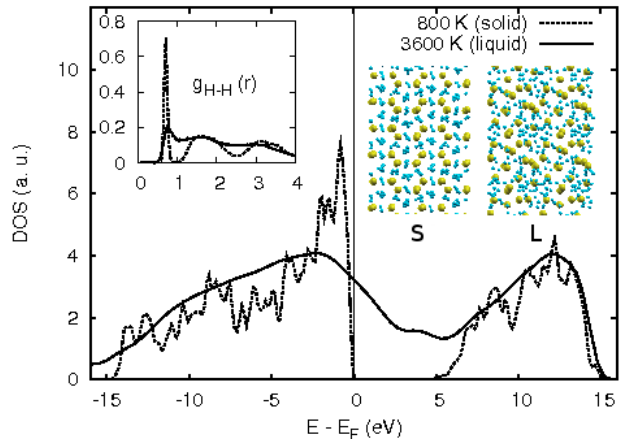}}%
\vspace{-0.25cm}
        \caption{DOS of Ar(H$_{2}$)$_{2}$ calculated at $P \sim 210$~GPa
                 ($V = 5.83$~\AA$^{3}$/particle) and different temperatures.
                 \emph{Inset}: Averaged radial H-H pair distribution 
                 function g$_{\rm H-H}$ obtained for the liquid (solid line) 
                 and solid (dashed line) phases (distance is in units of \AA). 
                 Snapshots of liquid (L) and solid (S) configurations as generated 
                 in the AIMD simulations; dissociated H$_{2}$ molecules 
                 are observed in L.}
\label{fig:DOS-HH}
\end{figure}

Results from our AIMD simulations are shown  
in Figure~\ref{fig:ub-melting} and can be summarized as follows:
(i)~the lattice of H$_{2}$ molecules
melts at temperatures significantly lower than the whole crystal does
so leading to mixtures of liquid H$_{2}$ and solid Ar over
wide $P - T$ ranges; 
(ii)~the fusion of the H$_{2}$ lattice occurs at temperatures  
very close to the melting line of pure hydrogen while Ar(H$_{2}$)$_{2}$ 
as a whole practically reproduces the melting behavior 
of pure Ar up to $\sim 60$~GPa; (iii)~the value of the estimated 
$\partial T_{m} / \partial P$ slope is negative within the pressure interval 
$60 \le P \le 110$~GPa whereas positive elsewhere.       
Results~(i) and~(ii) can be interpreted in terms of similar arguments than 
recently disclosed in solid mixtures of Ne-He and Ar-He, namely: 
lattices composed of same-species particles effectively
behave like not interacting one with another but following alike 
physical trends (bond distance, compressibility, etc.) than found in   
their respective pure system~[\onlinecite{cazorla09}].
In fact, we recently conjectured the superior energy stability 
of the MgCu$_{2}$ structure over MgZn$_{2}$
in Ar(H$_{2}$)$_{2}$, as rigorously demonstrated here, using this type 
of reasoning and crystal symmetry arguments~[\onlinecite{cazorla09}].  
Regarding result~(iii), this is in itself a manifestation of
a very peculiar and intriguing physical phenomenon. 
In general, systems presenting negative melting slope are 
characterized by open crystalline structure (water and graphite), 
surpassing promotion of valence electrons  
in the fluid phase (alkali metals) 
or continuous changing interparticle interactions
(molecular hydrogen). In the present case either open crystalline structure 
or electronic band promotion effects (energetically prohibitive) can be 
ruled out, so changes in the interparticle interactions holds as 
the likely cause. 
The question to be answered next then is:
what is the exact nature of these changing interactions?, or
more precisely, 
do they relate to dramatic changes in electronic structure?
are the ionic degrees of freedom and effective decoupling of the 
Ar and H$_{2}$ lattices crucial to them?
In order to detect possible electronic phase transitions   
and/or molecular dissociation processes, 
we performed meticulous averaged DOS and H-H radial 
correlation function (${\rm g_{H-H}}$) analysis over the atomic 
configurations generated in the AIMD runs. 
Within the thermodynamic range $60 \le P \le 110$~GPa and $0 \le T \le 3500$~K
(where $\partial T_{m} / \partial P \le 0$), we found that 
Ar(H$_{2}$)$_{2}$ is always an insulator material, either solid or liquid, 
where H$_{2}$ molecules remain stable.
Therefore, dramatic changes in electronic structure can be discarded as the 
originating mechanism behind the alternating sign of $\partial T_{m} / \partial P$. 
Actually, on the contrary case, one would have expected the value of 
$\partial T_{m} / \partial P$ to be negative also beyond $\sim 110$~GPa  
since in principle there is not reason to think that 
the originating electronic effects go missing at higher-$P$. 
Interestingly, at compressions above $\sim 210$~GPa and temperatures high enough 
for the system to melt we did observe closure of the electronic energy band
gap; according to ${\rm g_{H-H}}$ analysis and visual recreation of 
the AIMD configurations, this electronic phase transition is a 
consequence of emergent H$_{2}$ dissociation processes. 
In Figure~\ref{fig:DOS-HH}, we enclose DOS and ${\rm g_{H-H}}$ results that 
illustrate this insulator-to-metal phase transition. It is worth noticing 
that this result is consistent with shock-wave compression 
experiments~[\onlinecite{weir96}] and \emph{ab initio} investigations~[\onlinecite{bonev04}] 
performed on pure hydrogen.  
With regard to the predicted negative melting slope, 
ionic effects are the likely 
subjacent cause. Considering results~(i) and~(ii) above,  
one can envisage a plausible explanation of this phenomenon: 
since the lattice of H$_{2}$ molecules melts at temperatures much lower than the 
lattice of Ar atoms does, collisions between diffusive H$_{2}$ molecules and 
localized Ar atoms act like effective thermal-like excitations on 
the last that ultimately provoke the global melting of the system 
at temperatures below those of pure Ar. H$_{2}$ itself
presents negative melting slope beyond $\sim 60$~GPa so this effect  
is echoed, and enhanced significantly, in the melting line of 
Ar(H$_{2}$)$_{2}$ (see Figure~\ref{fig:ub-melting}).       
Beyond $\sim 110$~GPa, the stability of the Ar lattice becomes further 
reinforced so the effect of H$_{2}$ collisions there 
is just to deplete the slope of the global melting curve in comparison 
to that of pure Ar.
In order to test the validity of such hypothesis, we performed a series of
AIMD simulations at $P \sim 110$~GPa ($V = 7.50$~\AA$^{3}$/particle) in which 
we kept the lattice of H$_{2}$ molecules \emph{frozen} and left the lattice 
of Ar atoms to evolve. 
Under these conditions we found that fusion of the Ar structure occurs at 
temperatures $\sim 900$~K above that of pure Ar, hence the original
melting mechanism proposed seems to be corroborated. 
Analogous melting behavior than described here can be expected  
in other rare gas-H$_{2}$ compounds, like Xe(H$_{2}$)$_{7}$, and maybe also
in SiH$_{4}$(H$_{2}$)$_{2}$ and CH$_{4}$(H$_{2}$).

\section{Concluding Remarks}
\label{subsec:remarks}

To summarize, we have studied the behavior of Ar(H$_{2}$)$_{2}$ under 
pressure at low and high temperatures using computational first-principles 
techniques. As results, we have unravelled (i)~temperature(pressure)-induced solid-solid 
phase transitions that may resolve the existing discrepancies between the sets of 
R and IR experimental data, and (ii)~an anomalous melting phenomenon consisting 
of negative melting slope. The atypical melting line of Ar(H$_{2}$)$_{2}$ 
can be understood in terms of the decoupling of H$_{2}$ and
Ar ionic degrees of freedom and of coexistence of same-species liquid and solid 
phases. Metallization of liquid Ar(H$_{2}$)$_{2}$ is predicted at high $P - T$ 
conditions.

\acknowledgments

DE acknowledges support of MICINN of Spain (Grants 
No.~CSD2007-00045 and ~MAT2007-65990-C03-01).
The authors acknowledge computational resources on the U.K. National 
Supercomputing HECToR service.

\end{document}